\documentclass[12pt]{article}

\pdfoutput=1

\usepackage{color}
\usepackage{graphicx}

\usepackage{amssymb}
\usepackage{amsmath}

\newcommand{\rf}[1]{(\ref{#1})}
\newcommand{\beq}{\begin{equation}}
\newcommand{\eeq}{\end{equation}}
\newcommand{\bea}{\begin{eqnarray}}
\newcommand{\eea}{\end{eqnarray}}

\newcommand{\e}{\mbox{e}}
\renewcommand{\d}{\mbox{d}}
\newcommand{\g}{\gamma}

\newcommand{\lam}{\lambda}

\newcommand{\bt}{\beta}
\renewcommand{\a}{\alpha}

\newcommand{\m}{\mu}

\newcommand{\tht}{\theta}
\newcommand{\om}{\omega}
\newcommand{\del}{\delta}
\newcommand{\Del}{\Delta}
\newcommand{\sg}{\sigma}
\newcommand{\kp}{\kappa}
\newcommand{\vph}{\varphi}

\newcommand{\ra}{\rangle}
\newcommand{\la}{\langle}
\newcommand{\lla}{\left\la}
\newcommand{\rra}{\right\ra}

\newcommand{\cD}{{\cal D}}

\newcommand{\cO}{{\cal O}}

\newcommand{\tH}{{\tilde{H}}}

\newcommand{\hg}{{\hat{g}}}

\newcommand{\bg}{{\bar{g}}}
\newcommand{\bmu}{{\bar{\mu}}}

\newcommand{\no}{\nonumber}

\begin{document}

\begin{center}
\vspace{24pt}
{\Large \bf A note on the Lee-Yang singularity coupled to 2d quantum gravity}
\vspace{30pt}

{\sl J. Ambj\o rn}$\,^{a,b}$,
{\sl A. G\"{o}rlich}$\,^{a,c}$,
{\sl A.C. Ipsen}$\,^{a}$
and 
{\sl H.-G. Zhang}$\,^{c}$

\vspace{24pt}

{\footnotesize

$^a$~The Niels Bohr Institute, Copenhagen University\\
Blegdamsvej 17, DK-2100 Copenhagen \O , Denmark.\\
email: ambjorn@nbi.dk, goerlich@nbi.dk, asgercro@nbi.dk\\

\vspace{10pt}

$^b$~Radboud University Nijmegen\\
Institute for Mathematics, Astrophysics and Particle Physics (IMAPP),\\  
Heyendaalseweg 135, 6525 AJ Nijmegen, The Netherlands.\\

\vspace{10pt}

$^c$~Institute of Physics, Jagiellonian University,\\
Reymonta 4, PL 30-059 Krakow, Poland.\\
email: zhang@th.if.uj.edu.pl

}

\vspace{48pt}
\end{center}


\begin{center}
{\bf Abstract}
\end{center}

\noindent 
We show how to obtain the critical exponent of 
magnetization in the Lee-Yang edge singularity model coupled to 
two-dimensional quantum gravity.
  
\newpage

\section{Introduction}\label{introduction}

Two-dimensional quantum Liouville gravity and 
the theory of random triangulations (or matrix models) 
most likely describe the same theory, two-dimensional 
quantum gravity coupled to conformal field theories 
with a central charge $c\leq 1$. The two realizations
are sufficiently different that the ``proof'' that they 
describe the same theory is basically by 
comparing the result of calculations of certain 
``observables''. The major problem of such a comparison
has been to identify the observables to be compared in 
the two formulations. This problem has to a large 
extent been solved in \cite{mss} for one and two-point 
correlation functions and in \cite{bz} for three--
and four-points correlation functions. 
Here  we will address an observable, the 
so-called ``magnetization'' at the Lee-Yang edge
singularity. We will show how the general assumptions
of operator mixing put forward in \cite{cgm,mss,bz} 
allow us to obtain agreement between the 
critical exponent of the Lee-Yang ``magnetization''
calculated in quantum Liouville gravity and  using 
matrix models. 

The rest of this article is organized as follows:
in the next section we recapture how to calculate
the magnetization exponent $\sg$ in the Ising 
model and at the Lee-Yang edge singularity using 
standard conformal field theory. In sec.\  \ref{operatormix}
we then show how to reconcile Liouville and matrix
model results. 
 
\section{Ising models and dimer models}

The Ising model on an arbitrary connected graph $G_V$ with $V$ vertices
and $L$ links is defined by 
\beq\label{1.1}
Z_{G_V} (\beta,H) = \sum_{\{\sg_i\}} 
\exp \Big(\bt \sum_{\la ij\ra=1}^L \sg_i\sg_j +
H\sum_{i=1}^V \sg_i\Big),
\eeq
where the Ising spin $\sg_i$ (which can take values $\pm 1$) 
is located at vertex $i$, $\la ij\ra$  symbolizes
that vertices $i$ and $j$ are neighbors in $G_V$, and $\beta$ and $H$ 
signify inverse the temperature and a magnetic field, respectively.   

If $G_V$ is a regular two-dimensional lattice, e.g.\ a square lattice,
the partition function $Z_{G_V}(\beta,H=0)$ has a second order 
phase transition for a certain value $\beta_c$ in the limit $V\to \infty$.
Let us calculate 
\beq\label{1.2}
\lla \e^{H \sum_i \sg_i} \rra_{\bt=\bt_c,H=0} = \e^{-F_{G_V}(H)},
\eeq
using the partition function
$Z_{G_V}(\bt_c,0)$. For large $V$ the free energy $F_{G_V}(H)$ 
becomes extensive and the magnetization $m$ is given by
\beq\label{1.3}
F_{G_V}(H) = f(H) \,V\,(1+o(V)),~~~
m =-\frac{df}{dH} \sim |H|^{\sg}, ~~~\sg=\frac{1}{15},
\eeq
for small $H$.

The two-dimensional Ising model at its critical point $\bt_c$ is a conformal 
field theory with central charge $c=1/2$. Let us recall how 
the above result is derived using conformal field theory. Consider 
a conformal field theory and let $\Phi$ be a primary operator 
with scaling dimension $\Del_0$, i.e.\ $\Phi(\sqrt{\lam}\, x) 
= \lam^{-\Del_0}\Phi(x)$
(we consider $\Phi$ to be the product of its holomorphic and anti-holomorphic 
parts, i.e.\ real). Under a scaling $x\to \sqrt{\lam}\, x$ we thus have
\beq\label{1.4a}
A = \int d^2x \to \lam \, A,~~~~
D_0= \int d^2x \; \Phi(x) \to \lam^{1-\Del_0} D_0.
\eeq
We can study a ``deformation'' away 
from the conformal point by adding the term 
\beq\label{1.4c}
\del \;D_0=\del \int \d^2x \; \Phi,~~~~~~~[\del]= [A]^{\Del_0-1} 
\eeq
to the action. The last equation in \rf{1.4c} states
the dimension of the coupling constant $\del$ in terms 
of the dimension of the area $A$ of the 2d universe. 
As in eq.\ \rf{1.2} we can write
\beq\label{1.4d}
\lla e^{-\del D_0} \rra_0 = \e^{-F_A(\del)},
\eeq
where the average is calculated at the critical point. 
For large areas $A$ we expect $F_A$ to be extensive. For 
dimensional reasons we thus have, $\del$ being the only
coupling constant,
\beq\label{1.5}
F_A(\del) = f(\del) A (1+o(A)),~~~~f(\del) = k\, \del^{\frac{1}{1-\Del_0}}.
\eeq
The ``$\Phi$ magnetization'' is thus
\beq\label{1.6}
m_\Phi = -\frac{df}{d \del} \sim \del^{\Del_0/(1-\Del_0)},~~~
~~{\rm i.e.}~~\sg_{\Phi} = \frac{\Del_0}{1-\Del_0}.
\eeq
Applying this to the spin operator $\Phi_{1,2}$ of the 
(3,4) minimal conformal field theory which has central 
charge $c=1/2$ and  corresponds to the 
Ising model, we have $\Del_0 = 1/16$ and thus $\sigma_{\Phi_{1,2}}= 1/15$
in agreement with \rf{1.3}. For the (2,5) minimal conformal field theory
which has $c=-22/5$ 
there is only one non-trivial primary operator, again  $\Phi_{1,2}$, and 
$\Del_0= -1/5$. The corresponding magnetization exponent is  
$\sigma_{\Phi_{1,2}}= -1/6$.

Everything said above can be directly transferred to quantum Liouville
gravity as long as we consider the partition function for a 
fixed area which we then take large to avoid finite size effects. 
More precisely,  the partition function for a conformal field 
theory with central charge $c$ coupled to the Liouville field and with 
the area of the 2d ``universe'' fixed to be $A$ is defined as  
\beq\label{1.7}
Z_A = \int \cD \vph \cD \psi \;\e^{-S_L(\vph,\hg) -S_c(\psi,\hg)} \;
\del\left( \int d^2x \sqrt{\hg}\,\e^{\a \vph} -A\right).
\eeq
In \rf{1.7}  $S_c(\psi)$ is the matter action and $S_L(\vph)$ the Liouville 
action. $\hg_{ab}$ is a fiducial metric in the decomposition of the 
metric $g_{ab}=e^{\vph}\hg_{ab}$, thereby defining the Liouville field.
Changing variables $\vph \to \vph +\rho$ in the functional integral 
allow us to obtain (for surfaces with spherical topology)
\beq\label{1.8} 
Z_A \sim A^{\g_0 -3},~~~~\g_0= \frac{c-1-\sqrt{(25-c)(1-c)}}{12}.
\eeq

For a given conformal field theory and a given primary field $\Phi$,
the  observable $D_0$ defined above and the area $A$ are changed to 
\beq\label{1.8}
D = \int d^2x \sqrt{\hg} \; \e^{\bt \vph} \Phi, ~~~~~~~
A = \int d^2 x \sqrt{\hg} \; \e^{\a \vph} I
\eeq
In particular the area has become  an observable on equal footing 
with $D$, associated with the (trivial) 
primary field $I$ (the identity). The coefficients $\bt,\a$ are 
determined by the requirement that the observables $D$ and $A$ 
are invariant under diffeomorphisms and in 2d this 
implies that they are invariant under 
conformal transformations \cite{ddk}. 
However, $D$ still has a scaling dimension
relative to the area $A$. Let us define the expectation 
value of an observable $\cO$ for fixed area as  
\beq\label{1.9}
\la \cO\ra_A = \frac{1}{Z_A}\int \cD \vph \cD \psi \;\cO \; 
\e^{-S_L(\vph,\hg) -S_c(\psi,\hg)} \;
\del\left( \int d^2x \sqrt{\hg}\,\e^{\a \vph} -A\right).
\eeq  
One has 
\beq\label{1.10}
\la f(\lam^{-\bt/\a} D)\ra_{\lam A} = \la f(D) \ra_A
\eeq
for any function $f$. This follows by the change of 
integration variable $\vph \to \vph +\a^{-1}\log\lam$ in 
the functional integral \rf{1.9}. In particular we have 
\beq\label{1.11}
\la D\ra_{\lam A} = \lam^{\bt/\a} \la D \ra_A,~~~{\rm i.e.}~~~
1-\Del = \frac{\bt}{\a}, 
\eeq
by analogy with \rf{1.4a}. The scaling dimension $\Del$ is thus 
determined by $\a$ and $\bt$ and is given by the KPZ formula \cite{kpz} 
\beq\label{1.12}
\Del = \frac{ \sqrt{1-c +24 \Del_0} -\sqrt{1-c}}{\sqrt{25-c} -\sqrt{1-c}}.
\eeq

As in the ordinary conformal field theory case we can define 
the ``magnetization'' related to $\Phi$ by considering the 
perturbation away from the conformal point by the action 
\beq\label{1.12}
\del\, D = \del \int d^2x \,\sqrt{\hg} \; \e^{\bt \vph} \, \Phi,
~~~~~~[\del] = [A]^{\Del-1},
\eeq
in analogy with \rf{1.4c}. As in \rf{1.4d} we have 
\beq\label{1.13}
\lla \e^{-\del\,D}\rra_A = \e^{-F_A(\del)},~~~~
F(\del) = f(\del) A(1+G(A)),~~~~f(\del) = k \,\del^{1/(1-\Del)}.
\eeq
The ``magnetization'' is thus 
\beq\label{1.14}
m = -\frac{df}{d\del} \sim \del^{\Del/(1-\Del)},~~~{\rm i.e.}
~~~~\sg= \frac{\Del}{1-\Del}.
\eeq
In the case of the Ising model (i.e.\ $c=1/2$) coupled to the Liouville field
the exponent $\Del_0$ changes from 1/16 to 
$\Del=1/6$ according to \rf{1.12}. Thus we find that $\sg_0$ changes from
1/15 to $\sg=1/5$. This value was first obtained using the random 
matrix models in \cite{bk} and is a strong test of the equivalence 
between the continuum limit of the random surface models coupled 
to matter and quantum Liouville gravity. Applied to the  (2,5) minimal 
conformal field theory coupled to the Liouville field, $\sg_0$ changes 
from -1/6 to $\sg=-1/3$.

Finally it can be convenient to consider the grand partition function 
where the area is not kept fixed
\beq\label{1.14a}
Z(\mu,\del) = \int dA \; Z_A \;\e^{-\m A} \la \e^{-\del D} \ra_A \sim
\Big(\mu +k \del^{1/(1-\Del)}\Big)^{2-\g_0}.
\eeq
We obtain 
\beq\label{1.14b}
Z(\m,0) \sim \m^{2-\g_0},~~~~~Z(0,\del) = \del^{2-\g(\del)},~~~~
\g(\del) = \frac{\g_0-2\Del}{1-\Del}.
\eeq
We also observe that if the action $\m A +\del D$ is viewed as
a small perturbation away from the conformal point $\m=\del=0$
and $\m$ and $\del$ are of the same order of magnitude, 
the singular behavior of $Z(\mu,\del)$ is 
dominated by $\m^{(2-\g_0)}$ if the scaling dimension $\Del >0$.
If $\Del <0$, as can be the case for non-unitary conformal 
field theories, the singular behavior of $Z(\m,\del)$ will be dominated 
by $\del^{(2-\g(\del))}$. We note for future reference that 
for  the (2,5) minimal conformal field theory $\g_0= -3/2$ and 
$\g(\del)= -1/3$. In a grand canonical context it is 
natural to define 
\beq\label{1.14c}
Z(\m,\del)= \e^{-F(\m,\del)}, ~~~\la A\ra_{\m,\del} = -\frac{dF}{d\m},
~~~~M(\del) =  -\frac{dF}{d\del} = m(\del) \la A \ra_{\m,\del},
\eeq
and we have 
\beq\label{1.14d}
\la A\ra_{\m,\del} = \frac{1}{\m +k \del^{1/(1-\Del)}},
~~~~m(\del) \sim \del^{\Del/(1-\Del)}.
\eeq
For a given value of $\del$  we have 
\beq\label{1.14e} 
\la A \ra_{\m,\del} \to \infty ~~~~{\rm for}~~~\m \searrow \bar{\m}(\del),
\eeq
where the condition 
\beq\label{1.14f}
\bmu (\del) +k \del^{1/(1-\Del)} = 0
\eeq
determines the ``critical'' value of the cosmological constant $\m$
for a given value of $\del$. In particular we have 
\beq\label{1.14g}
\frac{d \bmu}{d \del} \sim m(\del).
\eeq

\subsection{Dimers}

Consider the Ising model on the graph $G_V$ defined above.
It has a   high temperature expansion
\bea\label{1.15} 
Z_{G_V} &=& (2\cosh H)^V (\cosh \bt)^L \times \\  
&&\Big[ 1+ \tanh^2H[\tht(1)\bt +O(\bt^2)] + 
\tanh^4H [\tht(2) \bt^2+0(\bt^4)]+\cdots \Big]\no
\eea
where $\tht(n)$ is the number of ways one can put down $n$ dimers
on the graph $G_V$ without the dimers touching each other (so-called
hard dimers). 
For imaginary magnetic fields it is thus possible to take the 
high temperature limit where $\bt \to 0$ and $H = i \tH \to i \pi/2$
in such a way that  $\xi =\bt\,  \tanh^2H$ is kept fixed. In this limit 
the terms in the bracket $[\cdots]$  in eq.\ \rf{1.15} become 
the partition function 
\beq\label{1.16}
Z_{G_V}(\xi) = \sum_n \tht(n) \xi^n,~~~~~~  (\xi =-\bt\,  \tan^2\tH)
\eeq
of a hard dimer model with fugacity $\xi$ (which is negative for 
$\tH \in ]0,\pi/2[$). For $\bt < \bt_c$ the Ising model is known to have
a phase transition at a critical, purely imaginary magnetic field
$H_c(\bt) = i \tH_c(\bt)$, the so-called Lee-Yang edge singularity
\cite{ly} (assuming as before that we have a regular graph $G_V$, 
and that  we take $V \to \infty$). It is also known that one can formally 
associate a ``magnetization'' to this transition \cite{fisher}:
\beq\label{1.17}
Z_{G_V}(\bt,\tH) = \e^{-F_{G_V}(\bt,\tH)}, ~~~~
F_{G_V}(\bt,\tH) \sim f(\bt,\tH)\, V,
\eeq
where  
\beq\label{1.18}
m(\bt) = -\frac{df}{d (\Del \tH)} \sim (\Del \tH)^{\sg_0},~~~~\Del \tH = 
\tH -\tH_c(\bt).
\eeq
The critical exponent $\sg_0$ is independent of $\bt$ for $\bt < \bt_c$. 
$\tH_c(\bt) \to \pi/2$ for $\bt \to 0$ and at this point we can 
extract $\sg$ from the dimer partition function \rf{1.16}. The dimer 
model has a critical point $\xi_c$ for a negative value of the fugacity $\xi$
which is precisely the limit of $-\bt \tan^2 \tH(\bt)$ for $\bt \to 0$.
Writing 
\beq\label{1.19}
Z_{G_V}(\xi) = \e^{-F_{G_V}(\xi)}, ~~~~F_{G_V}(\xi) = f(\xi) \,V,
\eeq
we obtain
\beq\label{1.20}
m = -\frac{df}{d\Del \xi} \sim (\Del \xi)^{\sg_0},~~~~\Del \xi = \xi-\xi_c.
\eeq
Finally it was shown in \cite{cardy} that the critical behavior 
of the Lee-Yang edge singularity or the hard dimer model could be 
associated with the (2,5) minimal conformal field theory, and from 
the above arguments, using conformal field theory
we know the corresponding $\sg_0 = -1/6$. This
is in agreement with numerical determinations of $\sg_0$ on regular 
lattices. 

Once this is established we can formally couple the Lee-Yang edge
singularity to quantum gravity in the sense that the critical behavior
is determined by the coupling between between the (2,5) conformal 
field theory and the Liouville theory. From the above we thus 
expect the magnetization exponent to change from -1/6 to -1/3,
and we would naively expect to obtain that result if we could 
explicitly solve the Ising model in an imaginary magnetic field
or the hard dimer model on the set of random graphs used to 
represent 2d gravity. In fact one can solve both models
on random graphs and one obtains $\sg = 1/2$ \cite{staudacher}.

 \section{Operator mixing}\label{operatormix} 
 
Let us for simplicity choose to work with the dimer model and discuss
how we can re-interpret the result of \cite{staudacher} using 
the general philosophy outlined in \cite{cgm,mss,bz}. The coupling of 
the dimer model to 2d gravity is done by summing over connected 
random graphs $G_V$. Here we restrict ourselves to a set of planar 
graphs, i.e.\ we define 
\beq\label{2.1}
Z_V(\xi) = \sum_{G_V} \frac{1}{C_{G_V}} \; Z_{G_V}(\xi),
\eeq
where $C_G$ denotes the order of the automorphism group of 
the graph $G$. We can introduce a grand partition function by 
also summing over graphs with different number of vertices:
\beq\label{2.2}
Z(g,\xi) = \sum_V g^V Z_V(\xi).
\eeq
Let us choose the simplest set of planar random graphs, namely the set 
where all vertices have order four. The corresponding $Z(g,\xi)$ 
can be calculated using matrix model techniques \cite{kazakov1,staudacher}.
For details we refer to \cite{staudacher}. Here we are only interested 
in the result. There exists a critical $\xi_c$. For each $\xi \geq \xi_c$
there exists a corresponding critical $\bg(\xi)$, the radius of 
convergence of the power series \rf{2.2}. We write
\beq\label{2.3}
Z_V(\xi) = \e^{-F_V(\xi)},~~~F_V(\xi)= f(\xi) V (1+o(V)),~~~~\
\log \,\bg(\xi) = f(\xi).
\eeq
On a regular lattice one  would clearly identify $f(\xi)$ as the 
free energy density and  expect to calculated the 
critical exponent $\sg$ according to \rf{1.20}. This calculation
was performed in \cite{staudacher}:
\beq\label{2.4a}
\bg(\xi) = \frac{1}{450\xi^2} \Big[(1+10\xi)^{3/2}-1\Big]-\frac{1}{30\xi}
\eeq
i.e.\ expanding around $\xi_c=1/9$ one obtains
\beq\label{2.4}
\Del\bg( \xi) +\frac{10}{9} \Del \xi=  \frac{20\sqrt{10}}{9} 
\,\Del \xi^{3/2} +O(\Del \xi^2),
\eeq
where 
\beq\label{2.4b}
\Del\xi = \xi-\xi_c,~~~~~~
\Del \bg(\xi)=\bg(\xi)-\bg(\xi_c).
\eeq
Differentiating \rf{2.4} after $\Del \xi$ we obtain
\beq\label{2.5}
\frac{ df}{d \xi}\Big|_{singular} = 
\frac{d \log \bg}{d\xi}\Big|_{singular} \sim \Del \xi^{1/2}.
\eeq
Clearly this is at odds with the KPZ value $\sg = -1/3$ mentioned 
above for the Lee-Yang edge singularity. We now explain how this is
due to operator mixing of $A$ and $D$, following the logic outlined 
in \cite{cgm,mss,bz}.

Denote $\bg(\xi_c)$ by $g_c$. The first observation is that 
\cite{kazakov1,staudacher} 
\beq\label{2.6}
Z(g,\xi_c)\Big|_{singular} = \Del g^{-1/3-2},~~~~\Del g= g_c-g,
\eeq
i.e.\ one obtains $\g(\del)$ (= -1/3) rather than $\g_0$ (=-3/2) 
for the critical susceptibility exponent related $Z$.
Naively one would have made the following identification in \rf{2.2}
\beq\label{2.7}
\left(\frac{g}{g_c}\right)^V \to \e^{-\m A}
\eeq
by introducing a scaling parameter $a$ (with the dimension of length 
relative to $A$ which we define to have the dimension of length squared) 
\beq\label{2.8}
\Del g = \m \,a^2,~~~~A = V \,a^2,~~~~a \to 0.
\eeq
But this is clearly too simple as it would imply a critical behavior
$\Del g^{-\g_0-2}$ in \rf{2.6} according to Liouville theory. 
$\Del g$ has to contain some reference to the coupling $\del$. In some 
sense this is natural since both $A$ and $D$ appear when we 
move away from the conformal point $\m=\del =0$. Fixing $\xi=\xi_c$ and 
changing $g_c \to g_c -\Del g$ is one way to move away from the 
point $g_c,\xi_c$ corresponding to $\m=\del =0$. The change \rf{2.4}
is another way, where we move along the critical line with 
a $\Del \bg(\xi)$ determined by $\Del \xi$. It should thus 
be compared to \rf{1.14f} where $\bmu(\del) +k \del^{1/(1-\Del)} = 0$, which 
defines ``criticality'' in the theory perturbed by the $A,D$ terms
in the action. This condition 
allows us to obtain the  relation between $\m\, a^2,\del \,a^3$ and 
$\Del g,\Del \xi$ if we, in acordance with \cite{cgm,mss,bz}, assume
that we deal with an analytic coupling constant redefinition. To 
lowest order, which is all we need, we thus have
\beq\label{2.9}
a^2 \,\mu = \Del g(\xi) + c_2 \Del \xi,~~~~
a^3 \del = c_3 \Del g(\xi)+\Del \xi.
\eeq
The condition $\bmu+ k\del^{2/3}=0$ implies
\beq\label{2.10}
\Del \bg(\xi) + c_3^{-1} \Del \xi = 
c_3^{-1}\Big(k^{-1}(c_3^{-1} -c_2)\Big)^{3/2}\,
\Del\xi^{3/2} + O(\Del \xi^2).
\eeq
Comparing with \rf{2.4} we obtain
\beq\label{2.11}
a^3 \,\del =\Del \xi + \frac{9}{10}\, \Del \bg(\xi),~~~
a^2\,\bmu(\del) = \Del \bg(\xi) + d\, \Del \xi,
\eeq
where $d= 10/9 -k (2 \sqrt{10})^{2/3}$.
This shows explicitly that $\Del g$ couples to $\del$ as 
anticipated from eq.\ \rf{2.6}. 

By construction we now 
have $\bmu (\del) \sim \del^{2/3}$ and thus the correct 
Liouville magnetization. Further, it is amusing to check 
how the ``wrong'' result \rf{2.5} actually becomes 
correct if one pays attention to the details 
\footnote{The author of \cite{staudacher} had no 
motivation to pay attention to details, since his work 
was done before the understanding of the possibility of operator 
mixing. In fact his seminal paper was precisely what eventually
led to this understanding.}. \rf{2.5} is obtained 
by differentiating \rf{2.4} after $\Del \xi$.
 For the special linear combination \rf{2.11}
eq. \rf{2.4} can  be written as
\beq\label{2.12}
a^3\,\del(\Del\xi,\Del \bg(\xi)) = 
\frac{20\sqrt{10}}{9} 
\,\Del \xi^{3/2} +O(\Del \xi^2),    
\eeq 
and differentiating with respect to $\Del \xi$ leads to 
\beq\label{2.13}
a^3 \frac{d  \del}{d \Del \xi} \sim \Del \xi^{1/2} ~~~
\mbox{or} ~~~\frac{d \del}{d \bmu} \sim \bmu^{1/2}+ O(a),
\eeq
i.e.\ according to eq.\ \rf{1.14g} exactly the correct Liouville 
equation for the magnetization $m$ if $\sg = -1/3$.

As mentioned one can also solve the Ising model coupled 
to 2d gravity \cite{kazakov,bk}.
The matrix models use the grand canonical ensemble of graphs,
i.e.\ starting with the partition function \rf{1.1} 
one  performs the same steps as 
in eqs.\ \rf{2.1} and \rf{2.2} for the dimer model.
 We thus have a partition function $Z(g,\bt,H)$.
Above the critical temperature we find a critical line with a critical
imaginary magnetic field \cite{staudacher} 
$H_c(\bt) = i \tH_c(\bt)$, $\bt < \bt_c$, 
analogous to what we find on a fixed graph.
For a fixed value of $\bt < \bt_c$ 
we have an equation similar to the dimer equation \rf{2.4} \cite{staudacher}
\beq
  \Del \bar g(\tH) +d_3 \Del \tH \sim  \Del  \tH^{3/2},~~~
\Del \tH = \tH -\tH_c(\bt),
\eeq
from which one would conclude that $\sigma  = 1/2$. As 
for the dimer model, this
should be understood as the result of operator mixing, 
and one should really write
\beq
a^2 \,\bmu = \Del \bg(\tH) +  d_2\Del \tilde H,~~~~
a^3 \del =  d_3^{-1}\Del \bg(\tH)+\Del \tilde H
\eeq
in order to recover the KPZ exponent.

Let us briefly mention the ordinary critical point of the Ising model
on a dynamical graph.
The critical exponents calculated in \cite{kazakov,bk} match the KPZ results,
even without accounting for mixing. Regarding $\sigma$ (and  $\gamma_{0}$)
one can explicitly check that the naive calculation is 
unaffected by operator mixing
(cf. the discussion after \eqref{1.14b}).
When the magnetic field is zero the model has a $\mathbb{Z}_2$ symmetry, 
which guarantees that the spin operator $\Phi_{1,2}$ is not 
turned on in the continuum language.
This, in turn, ensures that also the exponent $\alpha$ 
associated with the thermal operator 
$\Phi_{2,1}$ comes out ``right'' in \cite{bk}.

\section{Discussion}

We have shown how the calculation in \cite{staudacher} 
leads to agreement between the critical exponents of the   
``magnetization'' calculated in the hard dimer
model coupled to dynamical triangulations and in quantum Liouville
theory coupled to a (2,5) minimal conformal field theory.
The price of this agreement is that the naive separation between geometric and
matter degrees of freedom which might seem self-evident for
models of spins living on dynamical graphs can thus not be
taken for granted.

\subsection*{ Acknowledgments.} 
JA, AG and AI acknowledge support from the ERC Advanced Grant 291092
``Exploring the Quantum Universe'' (EQU) and by FNU, 
the Free Danish Research Council, through the grant 
``Quantum Gravity and the Role of Black Holes''. 
JA  were supported in part by Perimeter Institute of Theoretical Physics.
Research at Perimeter Institute is supported by the Government of Canada
through Industry Canada and by the Province of Ontario through the 
Ministry of Economic Development \& Innovation.


\begin{thebibliography}{99}

\bibitem{cgm}
  C.~Crnkovic, P.~H.~Ginsparg and G.~W.~Moore,
  Phys.\ Lett.\ B {\bf 237} (1990) 196.

\bibitem{mss}
  G.~W.~Moore, N.~Seiberg and M.~Staudacher,
  Nucl.\ Phys.\ B {\bf 362} (1991) 665.


\bibitem{bz}
  A.~A.~Belavin and A.~B.~Zamolodchikov,
  J.\ Phys.\ A {\bf 42} (2009) 304004
  [arXiv:0811.0450 [hep-th]].


\bibitem{staudacher}
  M.~Staudacher,
  Nucl.\ Phys.\ B {\bf 336} (1990) 349.


\bibitem{kazakov}
  V.~A.~Kazakov,
  Phys.\ Lett.\ A {\bf 119} (1986) 140.


\bibitem{bk}
  D.~V.~Boulatov and V.~A.~Kazakov,
  Phys.\ Lett.\ B {\bf 186} (1987) 379.

\bibitem{ly}
C.N. Yang and T.D. Lee, 
Phys . Rev . {\bf 87} (1952) 404.\\
T.D. Lee and C.N. Yang, Phys.\ Rev .\ {\bf 87} (1952) 410.

\bibitem{fisher}
M .E.\ Fisher, Phys.\ Rev.\ Lett.\ {bf 40} (1978) 1610. \\
P.J.\ Kortman and R.B.\ Griffiths, Phys.\ Rev.\ Lett.\ {\bf 27} (1971) 1439.

\bibitem{cardy}
J .L . Cardy, 
Phys.\ Rev.\ Lett.\ {\bf 54} (1985) 1354.

\bibitem{kpz}
  V.~G.~Knizhnik, A.~M.~Polyakov and A.~B.~Zamolodchikov,
  Mod.\ Phys.\ Lett.\ A {\bf 3} (1988) 819.


\bibitem{ddk}
  F.~David,
  Mod.\ Phys.\ Lett.\ A {\bf 3} (1988) 1651.\\
  J.~Distler and H.~Kawai,
  Nucl.\ Phys.\ B {\bf 321} (1989) 509.


\bibitem{kazakov1}
  V.~A.~Kazakov,
  Mod.\ Phys.\ Lett.\ A {\bf 4} (1989) 2125.




\end{thebibliography}
\end{document}